\documentclass[10pt, conference]{IEEEtran}
\usepackage[ruled,vlined]{algorithm2e}
\usepackage{cite,url}
\usepackage{amsmath,amssymb,amsfonts}
\usepackage[]{algorithm2e}
\usepackage{graphicx}
\usepackage{textcomp}
\usepackage{xcolor}
\usepackage{balance}
\usepackage{makecell}
\usepackage{ctable}
\usepackage{tabularx,booktabs}
\def\BibTeX{{\rm B\kern-.05em{\sc i\kern-.025em b}\kern-.08em
    T\kern-.1667em\lower.7ex\hbox{E}\kern-.125emX}}
\usepackage{multirow}
\usepackage{tabularx}
\makeatletter
\def\thickhline{%
  \noalign{\ifnum0=`}\fi\hrule \@height \thickarrayrulewidth \futurelet
   \reserved@a\@xthickhline}
\def\@xthickhline{\ifx\reserved@a\thickhline
               \vskip\doublerulesep
               \vskip-\thickarrayrulewidth
             \fi
      \ifnum0=`{\fi}}
\makeatother

\newlength{\thickarrayrulewidth}
\begin{document}
\title{An Overview of Channel Models for NGSO Satellites}
\author{\IEEEauthorblockN{Victor Monzon Baeza, Eva Lagunas, Hayder Al-Hraishawi, and Symeon Chatzinotas}
\IEEEauthorblockA{\textit{Interdisciplinary Centre for Security Reliability and Trust (SnT), University of Luxembourg, Luxembourg}\\
\textit{Email: \{victor.monzon, eva.lagunas, hayder.al-hraishawi, symeon.chatzinotas\}@uni.lu}} }
\maketitle
\begin{abstract}
Satellite communications industry is currently going through a rapid and profound transformation to adapt to the recent innovations and developments in the realm of non-geostationary orbit (NGSO) satellites. The growing popularity of NGSO systems, with cheap manufacturing and launching costs, has set to revolutionize the internet market. In this context, accurate channel characterization is crucial for the performance optimization and designing efficient NGSO communications, especially considering the dynamic propagation environment. While the Third Generation Partnership Project (3GPP) has provided some guidelines in Release 15, we observed certain divergence on the channel models considered in the literature, each with different assumptions and peculiarities. This paper provides an extensive review of the existing methods proposed for NGSO channel modeling that consider different orbits, frequency bands, user equipment, use-case and scenario peculiarities. The provided review discusses the channel modeling efforts from a contemporary perspective through trade-off analyses, classifications, and highlighting their advantages and pitfalls. The main goal is to provide a comprehensive overview of NGSO channel models to facilitate the selection of the most appropriate channel based on the scenario requirements to be evaluated and/or analysed. 
\end{abstract}

\section{Introduction}
\label{into}

Non-geostationary (NGSO) satellite constellations are considered a promising solution to provide global internet coverage in response to the emerging demand for global broadband, high speed, ultra-reliable, and low latency communications, driven primarily by the emergence of 5G networks \cite{SurveyGEO}. The Third Generation Partnership Project (3GPP) has studied the integration of satellites in 5G through Low Earth Orbit (LEO) constellations from Release-15 focusing on non-terrestrial network (NTN) channel models. Through this, several channel models were developed to support a range of deployment scenarios including urban, suburban, and rural mainly for S and Ka bands considering satellite links. While 3GPP provided valuable guidelines, they represent only a first step towards NGSO channel modeling which can be extended to characterize different propagation conditions and mobility from point of view of large- and small-scale parameters. Recently, megaconstellations such as OneWeb or Startlink offer new architecture for broadband services \cite{Architecture} in which the channel model is an open research issue. 

In the case of geostationary orbit (GEO) satellite, channel models have been proposed that have been validated for high frequencies. For instance, \cite{GEO1} uses channels that are deterministic and common across spot beams, characterized by signal to noise ratio (SNR). In \cite{GEO2}, a three-dimensional channel model is proposed for GEO operating at Q-band, including the path loss and shadowing  modelled by a Markov chain, and small-scale fading modelled by stochastic model. In addition, the movement of the receiver and the Rician factor influenced by environment scattering are considered. Nevertheless, these models are not valid when we have NGSO systems.
 
Several works in the literature have explored different alternatives for NGSO channel models. In \cite{Throughput}, the authors analyze and evaluate the throughput and capacity performance of LEO-based NTN for downlink without considering the communication channel model. They use average spectral efficiency to measure the channel serving a very small aperture terminal (VSAT). 

The inherent high-speed movement of satellites that characterizes LEO orbits in NGSO constellations requires special attention on the channel model. This feature will impose time-varying propagation conditions impacting the system performance in terms of mobility, availability, and user throughput. Therefore, in LEO satellite systems, it is fundamental to understand accurately the signal variations between mobile satellites and users on the ground. In this regard, \cite{5GLEO} evaluates the performance regarding handovers between LEO satellites, likewise, without channel model, assuming a fixed value for shadow and K-factor for fast fading according to a narrowband model defined by 3GPP in \cite{3GPP38811}. However, this assumption is not appropriate for broadband services expected by NGSO constellations. In \cite{5GLEO}, system-level simulations to evaluate the 5G New Radio mobility performance in LEO-based NTN was conducted, where a constant Line-of-sight (LOS) component was assumed for the satellite to user links due to the lack of a suitable channel model in terms of mobility. 
 
Following the trends from the terrestrial domain and in response to the spectrum congestion occurring in lower frequency bands, frequency bands such as Ku, K, Ka, and Q/V are becoming more and more popular for NGSO systems. Furthermore, future 6G deployment is expected to take place in similar high bands \cite{6G}, thus facilitating potential integration with NGSO systems. However, the signal propagation in these bands suffer severe tropospheric scintillation in the earth-satellite slant path where rain is the major source of attenuation. In addition, the channel shows extremely large Doppler frequency shift and Doppler spread, frequency dependence, extensive coverage range, and long communication distance which have not been incorporated into a complete channel model.

In this paper, we perform a detailed literature review of existing NGSO channel models, which evidence a lack of clear channel model for NGSO communication system evaluation. In this context, we have aimed to provide a classification of the available models, highlighting their main assumptions based on altitude, movement, path propagation model or statistical distribution. Finally, we conclude the paper with our vision of the next steps to be taken in terms of characterization of NGSO channel model and possible extensions of the 3GPP guidelines provided in Release 15.

The remainder of this paper is organized as follows: 
Section \ref{sec:model} presents a general NGSO model. In Section \ref{sec:review} the review of existing channel models is performed. Section \ref{sec:clas} shows the classification, tradeoff and open issues identified. Finally, in Section \ref{sec:conclusion}, the conclusions are summarized.

\section{NGSO System Overview} 
\label{sec:model}
\begin{figure}[!t]
\centering
\includegraphics[width = 0.5\textwidth]{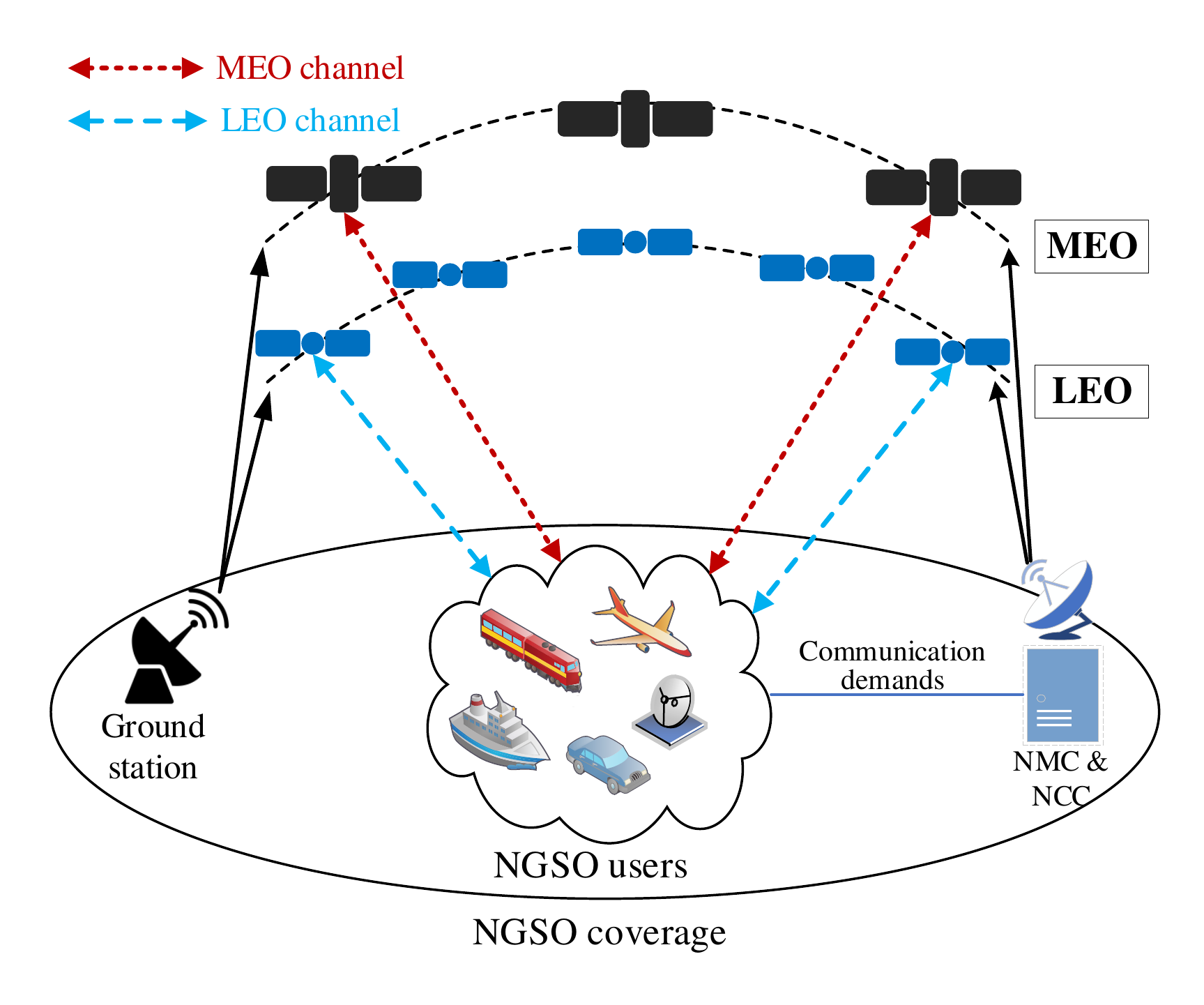}
\caption{A general schematic diagram of NGSO system.}
\label{fig:system_model}
\end{figure}

NGSO satellites on a geocentric orbit encompass the low Earth orbit (LEO), medium Earth orbit (MEO), which are orbiting continuously at lower altitudes than GEO satellites, which reduces their link losses and transmission latency. NGSO satellites typically cover smaller areas than the GEO satellites due to the lower orbits, and thus, NGSO systems are deployed as a constellation in order for achieving full Earth coverage. Particularly, an NGSO system consists of three main components: space segment, ground segment, and user segment (see Figure \ref{fig:system_model} for an illustration). The space segment is a constellation of NGSO satellites, while the ground segment is the gateway that relays data to and from the space segment, and the user segment can be a terminal with a transceiver and small antenna. Moreover, other essential entities are network management center (NMC) and network control center (NCC). The centralized NMC is the functional entity that is responsible of managing all the system aspects such as fault, configuration, performance, and security management. The NCC is the functional entity that provides real-time control signalling such as session/connection control, routing, and access control to satellite resources.

The frequency bands that can be used for NGSO satellites are regulated by the International Telecommunication Union (ITU) in order to ensure seamless coexistence with GEO satellite communications. Specifically, the frequency bands that can be shared for the NGSO-GEO scenario is summarized in Table \ref{table:ITU_regulations}. 
\begin{table}[!h] 
	\centering
	\caption{ITU regulations for the shared frequency bands between GEO and NGSO satellites.} \label{table:ITU_regulations}
	\begin{tabular}{|l|l|l|} 
		\hline
		Band                 & Frequency Range    & Priority of Operations\\
		\hline
		Ku				&  \makecell{ 10.7-10.95 GHz (space-to-Earth)\\
			11.2-11.45 GHz (space-to-Earth)\\
			12.75-13.25 GHz (Earth-to-space)}   & \makecell{GSO has priority\\ over NGSO\\	EPFD limits apply}
	\\
		\hline
		Ka & \makecell{17.8-18.6 GHz (space-to-Earth)\\
			19.7-20.2 GHz (space-to-Earth)\\
			27.5-28.6 GHz (Earth-to-space)\\
			29.5-30 GHz (Earth-to-space)} & \makecell{GSO has priority\\ over NGSO\\	EPFD limits apply}\\
		\hline
		Q/V & \makecell{37.5-42.5 GHz (space-to-Earth)\\
			47.2-50.2 GHz (Earth-to-space)\\
			50.4-51.4 GHz (Earth-to-space)}  & \makecell{Maximum degradation\\ of GSO reference links:\\ \textbullet\ Single entry (3\%) \\ \textbullet\ Aggregate (10\%)
		}\\
		\hline
	\end{tabular}
\end{table}

NGSO communication channels experience lower signal losses and propagation delays  comparing to the GEO satellites due to the lower altitudes. Thus, these advantages can be further utilized for miniaturizing user equipment, reducing users’ transmission powers, enhancing spectrum utilization, and targeting latency-critical applications. That would also allow NGSO users to adapt to commercial off-the-shelf modems, e.g. handheld and terrestrial IoT devices. Nevertheless, NGSO relative mobility causes variable receive power levels at user terminals, which is also affected by the ground antenna elevation and the slant path through the atmosphere. Besides, the well-known Doppler phenomenon and its impact on communication channels are inevitable in the NGSO systems. Specifically, Doppler effect creates time-varying frequency offsets and that will complicate the channel estimation process and increase the need for high channel estimation overheads. Motivated by these complications on NGSO channel modeling and the variety of link conditions and scenarios, next section will delve into analyzing the existing channel models in the open literature. 

\section{Review of existing channel models}
\label{sec:review}

In this section, we review and summarize the main characteristics of the channel models (CM) used in the literature so far. In this context, we have collected 13 different models for NGSO satellites where each one of them highlights a predominant channel feature.

\textbf{Model 1:}
In \cite{M1}, a CM for personal and mobile communications is initially proposed for constellations such as Globalstar and ICO. It is an elevation-dependent analog narrowband model for single-channel modeled by a two-state Markov chain: \begin{itemize}
    \item \textit{Good state}: corresponding to unshadowed areas with high received signal power or Line of Sight (LOS) component which is represented by Rician statistical distributions fading. 
    
    \item \textit{Bad state}: corresponding to areas with low received signal power also called non-LOS (NLOS) which is represented by Rayleigh or log-normal distribution fading.
\end{itemize}
The CM switches between both states, with no possibility of having LOS and NLOS components at the same time. In addition, this model defines the mean duration in each state depending on the type of environment, satellite elevation, and mobile user speed. On the other hand, it defines the correlation between two adjacent channels in the function of azimuth angle as well as the elevation to model the dual-channel diversity.

\textbf{Model 2:}
In \cite{M2}, a narrowband CM based on a Shadowed Rician Distribution (SRD) proposed in \cite{SRModel} is built. The large separation between the satellite and the ground user makes the signals from multiple onboard transmitters experience approximately the same propagation channel. Hence, this model assumes desired and interference signal powers are fully correlated rather than independent. Also, it is only applied for downlink (sky-to-ground) signals which are composed of a sum of the Squared Shadowed Rician (SSR) random variables modeling both LOS and NLOS propagation like conventional Rician fading, and also incorporates random shadowing into the LOS term to account for fluctuations due to the environment. This model is for LEO satellite operating in low frequencies.
   
\textbf{Model 3:}
In \cite{M3} a CM based on a stochastic geometry analysis is presented. This model assumes the desired signal experiences a Nakagami-\textit{m} fading, while the interfering signals propose using any fading model without an effect on analytical tractability. The operating frequency is assumed to be 2 GHz for massive LEO constellations. Different scenarios are considered by varying \textit{m}.

\textbf{Model 4:}
The Maseng-Bakken model (MBM) \cite{MBModel} has shown interesting properties to model rain attenuation for Ka and Q/V broadband satellite systems which has been subsequently adapted to give rise to new models. It is characterized by a time series synthesizer based on stochastic differential equations (SDE) to consider that rain attenuation follows a log-normal distribution. Time series synthesizers are also useful for system evaluation in case experimental rain attenuation time series are not available. In addition, this CM considers the rate of change of rain attenuation is proportional to the instantaneous value of rain attenuation. Later, International Telecommunication Union (ITU) proposed an updated version of MBM in \cite{ITUupdate}.
 
\textbf{Model 5:}
The CM considered in \cite{M5} only is based on the rain attenuation effects for both LEO and MEO constellations operating at Ka-band and above. This model proposes a novel time-series synthesizer for rain attenuation, incorporating the spatial correlation due to the rain on multiple channels. The model extends the MBM \cite{MBModel} using multi-dimensional SDE for the prediction of the joint exceeding probability of rain attenuation induced in LEO or MEO slant paths whose solution is used for the generation of time series of rain attenuation.

\textbf{Model 6:}
 In \cite{M6}, rain attenuation is modeled as a Weibull distribution. Then, the proposed CM is a time series synthesizer to generate said attenuation for satellite links operating at 10GHz and above. This model modifies MBM to consider the Weibull distribution resulting in a stochastic dynamic model based on the first-order SDE. Other works \cite{M6.1,M6.2} also describe the rain attenuation using Weibull distribution, where the prediction for rain attenuation in \cite{M6.1} is better than \cite{M6.2}.

\textbf{Model 7:}
In \cite{M8.1}, a Markov-based CM to determine the LOS state transitions was proposed as initial steps towards a model able to capture the time correlations. Later, in \cite{M8.2} the CM was completed with a model that predicts the signal strength variations caused by the effect of LOS obstructions considering stationary users on the LEO-to-Ground link. This model presented in \cite{M8.2} includes path loss, LOS changes, and the impact of the surrounding built-up structures in a suburban scenario using a low-complexity and intelligent ray-tracing approach that considers the general dependencies on the elevation angle.

\textbf{Model 8:} 
In \cite{3GPP38811}, 3GPP addressed the channel modeling for NTN from CM applied to TN. \cite{3GPP38811} 
studies propagation conditions adaptation. Despite the technical report considers model parameters for LEO satellites such as the probability LOS, clutter loss, and shadow fading, movement of the satellites is not supported.
Hence, the model does not define the correlations over time of the channel parameters. 3GPP groups in turn into several groups where each one considers a different parameter.    
 
\textbf{Model 9:}
For VLEO and LEO, in the report P.681 [5], the ITU defines its recommendations to predict the propagation impairments in the land mobile-satellite links. Even though the model includes some considerations regarding LEO satellites, only UE mobility is considered. An initial assessment based on extending the 3GPP Spatial Channel Model (SCM) to NTN is provided in [15]. However, further experimental channel campaigns are necessary to calibrate the model for NTN. For clutter loss  estimation can be estimated via ITU-R Recommendation P.2108 [17] and the UE needs to be outdoors in order to receive from a satellite.

\textbf{Model 10:}
In \cite{M12}, propagation phenomena for modeling channels in satellite communication systems operating at Ku, Ka, and V frequency bands are carried out by experimental work. Hence, we can extract the attenuation due to the tropospheric effects for a general CM in NGSO systems.
 
\textbf{Model 11:}
In \cite{M13} a CM is presented for very LEO satellites operating in millimeter waves (mmW), Ka, and Q based on experimental work. Regarding the user terminal, this needs to be outdoors in order to receive from a satellite. This model is interesting for the integration of satellites in 5G networks (Frequency Range 1 band in 5G when we consider mmW). However, this CM in \cite{M13} only consider the LOS channel because it employs highly directional antennas always pointing to the satellite. 

The measurement campaign considers several environments. In dense urban scenarios and low elevations, the LOS is blocked. This problem is solved in real satellite scenarios such as sub-urban or rural scenarios. Here, the shadowing effect shows a tendency to the log-normal standard deviation for the LOS component, complying with the 3GPP specification for channel modeling in \cite{3GPP38811}. In addition, this model makes use of recommendations in \cite{ITU618} for clutter loss estimation.

\textbf{Model 12:}
In \cite{M14} time series synthesizers and SDE are used again to generate a CM for cloud attenuation prediction. In this case, the methodology used in \cite{M14} for the estimation of cloud attenuation for GEO is extended to NGSO (MEO  and  LEO  cases) and modified to consider joint attenuation statistics (ILWC) of cloud attenuation for multiple paths and inclinations. The time series are generated from the Integrated Liquid Water Content model incorporating the time dependence of the elevation angle for NGSO channels. In \cite{M14.3} a unified space-time model for the estimation of cloud attenuation for Ka-band has been proposed also based on ILWC and SDE.

\textbf{Model 13:}
In \cite{M15}, the CM is defined to consider the dynamic behavior of the tropospheric scintillation and the movement of the LEO satellite. This model is generated by Weather Research and Forecasting software which simulates the space-time behavior of the turbulent troposphere. It is simulated for Ka-band and LEO satellites.

\section{Classification and tradeoff}
\label{sec:clas}
The first observation from Section III is that there is not a complet nor clear model for NGSO systems. In Table \ref{tab:comparison} we have compiled the first classification for the 13 models reviewed in the previous section. First of all, we differentiate between the models that are for broadband services and those that can be used for narrowband transmission, as is the case of IoT. We also differentiate between the technique used to obtain the model, that is, if the model is standardized, if it presents a statistical distribution, a geometric analysis or if it is exclusively dedicated to analyzing the attenuation. In this respect, the statistical models are more representative, being only two models that use a geometric analysis. These geometry models are the ones used for the case of having multipath. To do this, in addition to the CM we need to define the environment, which is not always evident.

Another classification block is the way the data is obtained. In this category, we include if the MC is obtained by simulation, if there has been an experimental measurement campaign or if there is a theoretical model. Most are obtained through simulations, evidencing the need for more channel characterization campaigns considering open air testing.

The available models are mostly presented for LEO satellites, with VLEO and MEO being less representative of NGSO within the modeling. Regarding the frequency bands, the Ka-band and low frequency are the most used. Considering the trend of ground segment stations moving towards Q/V bands, it would be beneficial to consider further evaluation and unification of channel models for such high-frequency bands.

As shown in Table \ref{tab:comparison}, most researchers mainly focus on using different distributions to represent the shadowing and small-scale fading and how the shadowing affects the LOS and scattered components. However, there is not a clear indication on when the multipath model needs to be considered. This is usually linked to non-directive antennas at user terminal side combined with urban-type of environment. In Table \ref{tab:los} a classification for each propagation component is performed. Classical distributions employed for terrestrial channels such as Rayleigh, Nakagami, or Rice are used in different current small-scale channel models, which are insufficient to take the propagation mechanism and the correlations among amplitude, angle, and Doppler frequency into account. Another popular method used in various models has been time series synthesizers using differential equations. However, the channel models used in this case do not differentiate between component LOS and NLOS. Markov chain models have also been used mainly for the LOS propagation component. We also highlight that there are models that have used a description based on one of the statistical distributions but without specifying which kind of view the propagation component would have. Furthermore, in the case of models 11 and 13, the CM is also described without specifying a known distribution. The ratio of LOS to NLOS components was also not considered in any CM.

As seen in the first classification, more than half of the models are dedicated exclusively to representing the channel based on the attenuation presented by the link. In Table \ref{tab:att} we have classified the different models for the types of attenuation they have considered. We can see how the two main phenomena are rain as the majority followed by clouds. Models 2, 10, and 13 have not specified exactly which phenomenon would be more predominant and have carried out the analysis in a generic way for the troposphere. Obviously, the attenuation impairments have to be defined based on the frequency of operation. 

Statistical distributions are once again an option to model this type of attenuation, in this case, the most representative is the log/normal distribution. Also, time series synthesizers are selected as an option to model rain, while cloud is modeled by a specific model which is ILWC. Model 13 has used software to perform simulations that represent the attenuation produced in the troposphere, while model 10 does not present any distribution, the results being obtained by experimental methods.

In the case of NGSO, the elevation angle is important since it has an effect on the system performance, whereas only models 1, 10, and 11 have considered elevation in their results. However, they have not included this angle in the model distribution. Also, model 1 has taken into account the azimuth on the correlation between channels.

Regarding the effect of the movement of the satellite or the user terminal, only models 1 and 5 have considered this effect, including Doppler in the model.

\begin{table*}[!t]
	\centering
	\caption{NGSO Channel Models Classification}
	\label{tab:comparison}
	\begin{tabular}{@{}l|l|l|l|l|l|l|l|l|l|l|l|l|l|@{}}
	
		\midrule \label{table1}
		\hspace{2.1mm} {Models}  &  
		\rotatebox[origin=c]{0}{\shortstack{M1}}&
	    \rotatebox[origin=c]{0}{\shortstack{M2}}&
	    \rotatebox[origin=c]{0}{\shortstack{M3}}&
		\rotatebox[origin=c]{0}{\shortstack{M4}}&
	    \rotatebox[origin=c]{0}{\shortstack{M5}}&
	    \rotatebox[origin=c]{0}{\shortstack{M6}}& 
		\rotatebox[origin=c]{0}{\shortstack{M7}}&
	    \rotatebox[origin=c]{0}{\shortstack{M8}}&
	    \rotatebox[origin=c]{0}{\shortstack{M9}}& 
		\rotatebox[origin=c]{0}{\shortstack{M10}}&
	    \rotatebox[origin=c]{0}{\shortstack{M11}}&
	    \rotatebox[origin=c]{0}{\shortstack{M12}}&
	    \rotatebox[origin=c]{0}{\shortstack{M13}}\\ \midrule
	    \multicolumn{1}{l|} {Narrowband}                              & \checkmark & \checkmark &            & \checkmark &            &            &            & \checkmark & \checkmark & \checkmark &            &            &            \\ \midrule
	    \multicolumn{1}{l|} {Broadband}                               &            &            & \checkmark &            & \checkmark & \checkmark & \checkmark & \checkmark & \checkmark &            & \checkmark & \checkmark & \checkmark \\\midrule 
	    \multicolumn{1}{l|} {Standardization}			               &            &            &            &            &            &            &            & \checkmark & \checkmark &            &            &            &           \\ \midrule
 		\multicolumn{1}{l|} {Statistical distribution }               & \checkmark & \checkmark & \checkmark &            &            & \checkmark & \checkmark &            &            &            &            &            &            \\ \midrule
 		\multicolumn{1}{l|} {Geometric analysis}                     &            &            & \checkmark &            &            &            & \checkmark &            &            &            &            &            &            \\ \midrule
 		\multicolumn{1}{l|} {Link Attenuation}                        &            &            &            & \checkmark & \checkmark & \checkmark &            &            &            & \checkmark & \checkmark & \checkmark & \checkmark \\\midrule 
 		\multicolumn{1}{l|} {Simulation Results}                      & \checkmark &            &            &            &            &            &            &            &            &            &            &            &            \\ \midrule
 		\multicolumn{1}{l|} {Experimental Results}                    &            &            &            &            &            &            &            &            &            &            & \checkmark &            &            \\ \midrule
 		\multicolumn{1}{l|} {VLEO satellites}                         &            &            &            &            &            &            &            &            & \checkmark &            & \checkmark &            &            \\ \midrule
 		\multicolumn{1}{l|} {LEO satellites}                          & \checkmark & \checkmark & \checkmark & \checkmark & \checkmark &            & \checkmark & \checkmark & \checkmark &            &            & \checkmark & \checkmark \\ \midrule
 		\multicolumn{1}{l|} {MEO satellites}                          & \checkmark &            &            &            & \checkmark &            &            &            &            &            &            & \checkmark &            \\ \midrule 
 		\multicolumn{1}{l|} {Low frequency band}                      & \checkmark & \checkmark & \checkmark &            & \checkmark &            &            & \checkmark & \checkmark &            &            &            &            \\ \midrule
 		\multicolumn{1}{l|} {Ku frequency band}                       &            &            &            & \checkmark &            & \checkmark &            &            &            & \checkmark & \checkmark &            &            \\ \midrule
 		\multicolumn{1}{l|} {Ka frequency band}                       &            &            &            & \checkmark &            & \checkmark &            &            &            & \checkmark & \checkmark & \checkmark & \checkmark \\ \midrule
 		\multicolumn{1}{l|} {Q/V frequency band}                      &            &            &            & \checkmark &            & \checkmark &            &            &            & \checkmark & \checkmark &            &            \\ 
		\bottomrule
	\end{tabular}
\end{table*}

\begin{table*}[!t]
	\centering
	\caption{LOS/NLOS Components}
	\label{tab:los}
	\begin{tabular}{@{}l|llll|lll|lllll|@{}}
\cmidrule(l){2-13}
                                                         & \multicolumn{4}{l|}{LOS Component}                                                                                                                                           & \multicolumn{3}{l|}{NLOS Component}                                                                                         & \multicolumn{5}{l|}{Non Specified}                                                                                                                                                                                            \\ \midrule
\multicolumn{1}{|l|}{Statistical Distribution / Models}    & \multicolumn{1}{l|}{M1}                        & \multicolumn{1}{l|}{M2}                        & \multicolumn{1}{l|}{M7}                        & M11                       & \multicolumn{1}{l|}{M1}                        & \multicolumn{1}{l|}{M2}                        & M3                        & \multicolumn{1}{l|}{M4}                        & \multicolumn{1}{l|}{M5}                        & \multicolumn{1}{l|}{M6}                        & \multicolumn{1}{l|}{M12}                       & M13                       \\ \midrule
\multicolumn{1}{|l|}{Rice Distribution}                  & \multicolumn{1}{l|}{\checkmark} & \multicolumn{1}{l|}{}                          & \multicolumn{1}{l|}{}                          &                           & \multicolumn{1}{l|}{}                          & \multicolumn{1}{l|}{}                          &                           & \multicolumn{1}{l|}{}                          & \multicolumn{1}{l|}{}                          & \multicolumn{1}{l|}{}                          & \multicolumn{1}{l|}{}                          &                           \\ \midrule
\multicolumn{1}{|l|}{Shadowed Rician Distribution (SRD)} & \multicolumn{1}{l|}{}                          & \multicolumn{1}{l|}{\checkmark} & \multicolumn{1}{l|}{}                          &                           & \multicolumn{1}{l|}{}                          & \multicolumn{1}{l|}{\checkmark} &                           & \multicolumn{1}{l|}{}                          & \multicolumn{1}{l|}{}                          & \multicolumn{1}{l|}{}                          & \multicolumn{1}{l|}{}                          &                           \\ \midrule
\multicolumn{1}{|l|}{Rayleigh Distribution}              & \multicolumn{1}{l|}{}                          & \multicolumn{1}{l|}{}                          & \multicolumn{1}{l|}{}                          &                           & \multicolumn{1}{l|}{\checkmark} & \multicolumn{1}{l|}{}                          &                           & \multicolumn{1}{l|}{}                          & \multicolumn{1}{l|}{}                          & \multicolumn{1}{l|}{}                          & \multicolumn{1}{l|}{}                          &                           \\ \midrule
\multicolumn{1}{|l|}{Log-normal Distribution}            & \multicolumn{1}{l|}{}                          & \multicolumn{1}{l|}{}                          & \multicolumn{1}{l|}{}                          &                           & \multicolumn{1}{l|}{\checkmark} & \multicolumn{1}{l|}{}                          &                           & \multicolumn{1}{l|}{}                          & \multicolumn{1}{l|}{}                          & \multicolumn{1}{l|}{}                          & \multicolumn{1}{l|}{}                          &                           \\ \midrule
\multicolumn{1}{|l|}{Weibull Distribution}               & \multicolumn{1}{l|}{}                          & \multicolumn{1}{l|}{}                          & \multicolumn{1}{l|}{}                          &                           & \multicolumn{1}{l|}{}                          & \multicolumn{1}{l|}{}                          &                           & \multicolumn{1}{l|}{}                          & \multicolumn{1}{l|}{}                          & \multicolumn{1}{l|}{\checkmark} & \multicolumn{1}{l|}{}                          &                           \\ \midrule
\multicolumn{1}{|l|}{Nakagami-m}                         & \multicolumn{1}{l|}{}                          & \multicolumn{1}{l|}{}                          & \multicolumn{1}{l|}{}                          &                           & \multicolumn{1}{l|}{}                          & \multicolumn{1}{l|}{}                          & \checkmark & \multicolumn{1}{l|}{}                          & \multicolumn{1}{l|}{}                          & \multicolumn{1}{l|}{}                          & \multicolumn{1}{l|}{}                          &                           \\ \midrule
\multicolumn{1}{|l|}{Time Series Synthesizer}            & \multicolumn{1}{l|}{}                          & \multicolumn{1}{l|}{}                          & \multicolumn{1}{l|}{}                          &                           & \multicolumn{1}{l|}{}                          & \multicolumn{1}{l|}{}                          &                           & \multicolumn{1}{l|}{\checkmark} & \multicolumn{1}{l|}{\checkmark} & \multicolumn{1}{l|}{}                          & \multicolumn{1}{l|}{\checkmark} &                           \\ \midrule
\multicolumn{1}{|l|}{Markov Model}                       & \multicolumn{1}{l|}{}                          & \multicolumn{1}{l|}{}                          & \multicolumn{1}{l|}{\checkmark} &                           & \multicolumn{1}{l|}{}                          & \multicolumn{1}{l|}{}                          &                           & \multicolumn{1}{l|}{}                          & \multicolumn{1}{l|}{}                          & \multicolumn{1}{l|}{}                          & \multicolumn{1}{l|}{}                          &                           \\ \midrule
\multicolumn{1}{|l|}{Non Specified}                      & \multicolumn{1}{l|}{}                          & \multicolumn{1}{l|}{}                          & \multicolumn{1}{l|}{}                          & \checkmark & \multicolumn{1}{l|}{}                          & \multicolumn{1}{l|}{}                          &                           & \multicolumn{1}{l|}{}                          & \multicolumn{1}{l|}{}                          & \multicolumn{1}{l|}{}                          & \multicolumn{1}{l|}{}                          & \checkmark \\ \bottomrule
\end{tabular}
\end{table*}

\begin{table*}[!t]
	\centering
	\caption{Attenuation}
	\label{tab:att}
	\begin{tabular}{@{}l|llll|l|lll|@{}}
\cmidrule(l){2-9}
                                                   & \multicolumn{4}{l|}{Rain}                                                         & Cloud & \multicolumn{3}{l|}{General Troposphere Effect}          \\ \midrule
\multicolumn{1}{|l|}{Model}                        & \multicolumn{1}{l|}{M4} & \multicolumn{1}{l|}{M5} & \multicolumn{1}{l|}{M6} & M11 & M12   & \multicolumn{1}{l|}{M2} & \multicolumn{1}{l|}{M10} & M13 \\ \midrule
\multicolumn{1}{|l|}{Log-Normal Distribution}      & \multicolumn{1}{l|}{\checkmark}  & \multicolumn{1}{l|}{}   & \multicolumn{1}{l|}{}   & \checkmark   &       & \multicolumn{1}{l|}{\checkmark}  & \multicolumn{1}{l|}{}    &     \\ \midrule
\multicolumn{1}{|l|}{Time Series Synthesizer}      & \multicolumn{1}{l|}{\checkmark}  & \multicolumn{1}{l|}{\checkmark}  & \multicolumn{1}{l|}{}   &     &       & \multicolumn{1}{l|}{}   & \multicolumn{1}{l|}{}    &     \\ \midrule
\multicolumn{1}{|l|}{Weibull Distribution}         & \multicolumn{1}{l|}{}   & \multicolumn{1}{l|}{}   & \multicolumn{1}{l|}{\checkmark}  &     &       & \multicolumn{1}{l|}{}   & \multicolumn{1}{l|}{}    &     \\ \midrule
\multicolumn{1}{|l|}{ILWC}                         & \multicolumn{1}{l|}{}   & \multicolumn{1}{l|}{}   & \multicolumn{1}{l|}{}   &     & \checkmark    & \multicolumn{1}{l|}{}   & \multicolumn{1}{l|}{}    & \checkmark   \\ \midrule
\multicolumn{1}{|l|}{SW Simulation:}               & \multicolumn{1}{l|}{}   & \multicolumn{1}{l|}{}   & \multicolumn{1}{l|}{}   &     &       & \multicolumn{1}{l|}{}   & \multicolumn{1}{l|}{\checkmark}   &     \\ \bottomrule
\end{tabular}
\end{table*}
 
\section{Conclusions and Discussion}
\label{sec:conclusion}
This paper provides an overview of the main channel models for NGSO constellations proposed in the open literature. A detailed classification has been performed to facilitate the comparison of the different channel models. We have evidence that there is no consensus on the NGSO model to be considered, as there is no complete model collecting all the necessary parameters to accurately characterize the challenges of NGSO channels. Nevertheless, we believe that the overview and comparison presented herein would be definitely beneficial for both academic and industrial community working on characterization and evaluation of NGSO systems.

Considering the observations highglighted in Section IV, we gather the following open research points:

\begin{itemize}
    \item The lack of consensus amoung the community regarding NGSO channel model is resulting in a proliferation of different works, each assuming different channel models. The latter poses significant challenges for comparison and validation of research outcomes. 
    \item While S band and Ka band seem to be well documented, there seems to be less information on channel models for high frequencies such as Q/V band, which is a key spectrum for the feeder link of NGSO constellations.
    \item There is no clear guidelines on when NLoS has to be assumed and how to statistically model such propagation environment when there is lack of knowledge of the specific surrounding environment.
    \item 3GPP Release 15 provides a first step towards unification of models for testing NGSO systems, particularly for NTN-TN integration. However, these are focused on S and Ka band, and generally simplistic models which do not consider e.g. performance loss due to Doppler imperfect compensation or impact of elevation angle into radiation patterns.

    
\end{itemize}

\section*{Acknowledgment}
This work has been supported by the Luxembourg National Research Fund (FNR) under the project MegaLEO (C20/IS/14767486). 

\balance
\bibliographystyle{IEEEtran}


\end{document}